# A large-scale real-life crowd steering experiment via arrow-like stimuli

Alessandro Corbetta[1,*], Werner Kroneman[1], Maurice Donners[2], Antal Haans[3],
Philip Ross[4], Marius Trouwborst[2], Sander Van de Wijdeven[2], Martijn Hultermans[2],
Dragan Sekulovski[2], Fedosja van der Heijden[2], Sjoerd Mentink[2], Federico Toschi[1]

[1]Department of Applied Physics, Eindhoven University of Technology, NL
[2]Philips Lighting, Eindhoven, NL
[3]Human Technology Interaction, Eindhoven University of Technology, NL
[4]Studio Philip Ross, Eindhoven, NL
[*]corresponding author. E-mail: a.corbetta@tue.nl

**Abstract -** We introduce "Moving Light": an unprecedented real-life crowd steering experiment that involved about 140.000 participants among the visitors of the Glow 2017 Light Festival (Eindhoven, NL). Moving Light targets one outstanding question of paramount societal and technological importance: "can we seamlessly and systematically influence routing decisions in pedestrian crowds?" Establishing effective crowd steering methods is extremely relevant in the context of crowd management, e.g. when it comes to keeping floor usage within safety limits (e.g. during public events with high attendance) or at designated comfort levels (e.g. in leisure areas).
In the Moving Light setup, visitors walking in a corridor face a choice between two symmetric exits defined by a large central obstacle. Stimuli, such as arrows, alternate at random and perturb the symmetry of the environment to bias choices. While visitors move in the experiment, they are tracked with high space and time resolution, such that the efficiency of each stimulus at steering individual routing decisions can be accurately evaluated a posteriori.
In this contribution, we first describe the measurement concept in the Moving Light experiment and then we investigate quantitatively the steering capability of arrow indications.

## 1. Introduction

Developing effective management strategies for the motion of pedestrian crowds is a compelling issue in the course towards highest safety and comfort standards in civil infrastructures. Crowd management involves routing pedestrian flows to ensure designated Level-of-Services [1]. This includes an ample spectrum of scenarios in terms, e.g., of crowd density, whose extreme cases are the prevention of dangerous overcrowdings in public gatherings/trafficked hubs (stations, stadiums, etc.) and the establishment of comfortable and uniform floor usage in leisure locations (museums, commercial areas, etc.).
Crowd steering is generally relegated to on-site stewards that, depending on necessities, sort the flow or route it via indications or - in extreme conditions - via physical barriers. Automatizing such steering procedures could be greatly beneficial: actions can be triggered in absence of or with less human supervision, the request for on-site manpower can be diminished, and the spatial granularity of guiding mechanisms can be increased, e.g. for fine-scale floor usage optimizations.
Automating steering is about influencing individual route choices which, in turn, depend on available information about alternative directions. Apart from the geometric characteristics of the environment, also the presence of signage and the relative density of the surrounding crowd plays a role. Currently, experiments comparing these factors have mostly been performed in virtual environments (e.g. [2,3,4]) and with single individuals or at low crowd densities. As of today, empirical evidence on their impact on the crowd movement as a whole is scarce and shows mixed and often contradictory results [5]. Quantitative real-life analyses of the effectiveness of visual stimuli at influencing individual routing decisions are thus a must toward automated steering. Comparing with a wide variety of crowding conditions (e.g., density levels) as well as ensuring high statistical resolution in the measurements are furthermore paramount, given the high variability in pedestrian behaviour [6].

We introduce here "Moving Light", a real-life experiment in which we targeted, in quantitative terms, a prototypical case of automated crowd steering: swaying – by means of visual stimuli – the route choice of

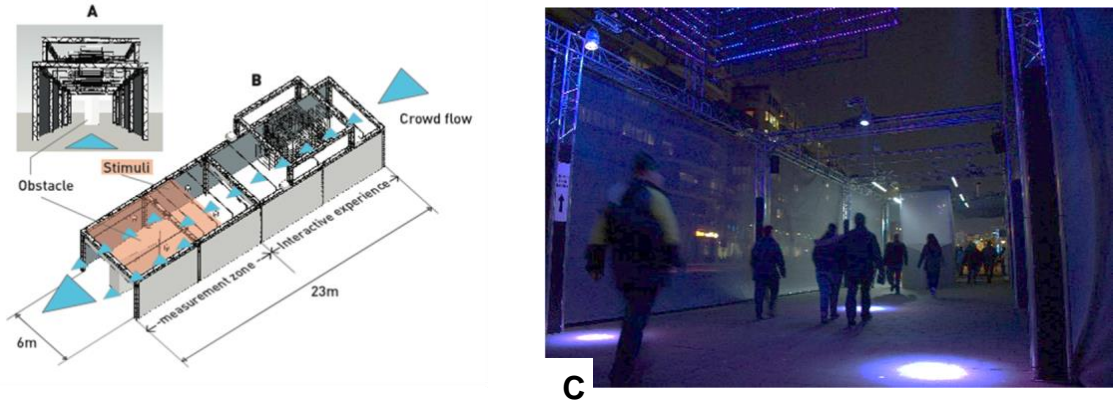

Figure 1: The "Moving Light" real-life pedestrian steer-and-track experiment, held by us as an exhibit at the Glow Festival in November 2017 (Eindhoven, The Netherlands, cf. [7]). (A) Sketch from frontal view as seen by an entering visitor; (B) aerial view - the light blue triangles indicate the direction of the crowd flow. (C) Picture of the exhibit in action (view from the right corner of the entrance).

The scientific core of the facility lays in its second half, the "measurement zone", where the visitors faced the decision to pass either on the left or on the right side of the central obstacle in order to leave the exhibit. We swayed this decision – otherwise expected left-right symmetric – with stimuli randomly changing every 3 minutes (cf. Figure 2). The measurement zone hosted two high-resolution pedestrian tracking systems for quantifying the steering effectiveness of the stimuli. Visitors entered the measurement zone after an "interactive experience" lasting 1 minute. This ensured a regular scheduling in the crowd flow and aimed at having quasi-independent visitors batches (i.e. experiments) proceeding to the measurement zone every minute.

pedestrians between two symmetric exits (Figure 1).

The experiment took place as one of the exhibits in the week-long 2017 Eindhoven Glow Light Festival [7]. The Glow Festival, occurring every year and running in the evenings, involves a city-wide uni-directional route in which visitors – in the order of hundreds of thousands – walk through exhibits related to illumination design and light art. In 2017, the event took place in the period $11^{th}$ - $18^{th}$ November. The unique experimental nature and the high attendance makes the festival a perfect location for analysing crowd dynamics.

In the Moving Light experiment, for the entire duration of the event, we subjected the visitors stream to stimuli which periodically changed every three minutes and which were chosen randomly from a pool of

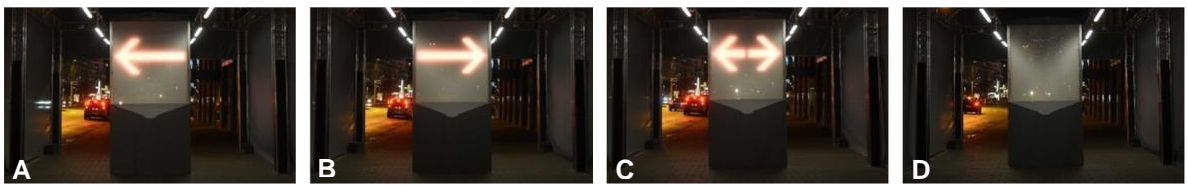

Figure 2: Arrow based stimuli (A, B) and related symmetric control conditions (C, D) employed to sway individual decisions for either exits and thus to steer the crowd. Each stimulus appeared at random on a LED matrix on the front side of the obstacle for intervals of three minutes. (A,B) arrows pointing to the left or the right aimed at steering pedestrians to, respectively, the exist on the left or on the right side of the obstacle. (C, D) Neutral control conditions encompassing, respectively, a doubly-sided arrow and absence of signage (no arrow displayed). In the following, these four stimuli are indicated in symbols for brevity respectively as " $\Leftarrow$ ", " $\Rightarrow$ ", " $\Leftrightarrow$ " and " $\diamond$ ".

18 (one stimulus appeared at once, stimuli were e.g. based on arrow indications or on illumination). About 140.000 of the festival visitors crossed our installation: we tracked them individually at high space and time resolution aiming at quantifying the impact of the stimuli on their individual trajectories and on their final exit choice. Visitors were not aware, or at least not notified, that trajectories were recorded and thus we can assume that no bias originated from the experimental setting (note that no feature, visual or otherwise, allowing personal identification was employed or stored in the process).

In this manuscript we analyse the steering performance of arrow stimuli as shown in Figure 2 (these form a subset of 4 stimuli in the pool of 18). Specifically, this contribution is structured as follows: in Section 2 we describe the Moving Light experimental setup. This includes the crowd steering hypothesis testing rationale and the tracking technologies employed. Then, in Section 3, we focus on the dynamics triggered by arrow stimuli and quantify their effectiveness and action range. Section 4 contains a closing discussion.

## 2. The Moving Light real-life experiment

The Moving Light installation, placed on a wide sidewalk, is framed as a 23m x 6m corridor

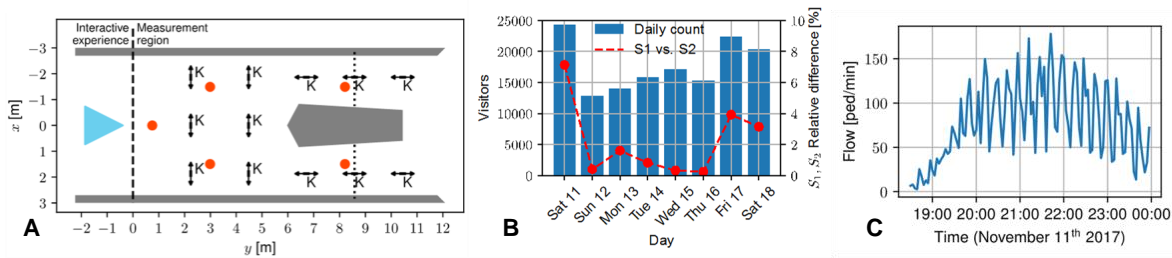

Figure 3: (A) Measurement systems at the Moving Light experiment. Pedestrian tracking has been operated via a Microsoft Kinect™ based system (S1, grid of 12 units, indicated with a "K" and with major Field-of-View angle aligned with the arrows) and via a Xovis AG commercial system (S2, sensor locations indicated by orange dots). Both system allowed full coverage of the measurement zone (cf. Figure 4A for a depth field image by the Kinect system grid). (B) Visitors count of the Moving Light throughout the 8 evenings of the festival. Bars: average count between the two measurement systems: Kinect-based (S1) and Xovis-based (S2). Dashed line: absolute difference among the counts of the two systems relative to the mean. Visitors peaked on Saturdays and Friday evenings. In these cases, maximum discrepancies in the counts (about 7%) are measured. (C) Visitors flow (ped./min) during the first evening (readings on windows 3 minutes long). A fluctuation with a period of about 15 minutes can be observed. This is due to a periodic light show located a few hundred meters upstream with respect to our installation.

(demarcated on either side by 3 meters high semi-transparent fabric fences), and it is formed of two adjacent zones, namely the "interactive experience" zone and the "measurement zone". These are consecutive along the visit and each one measures 11.5 m x 6m (i.e. half of the full length). The "interactive experience" comes first. It encloses the artistic contribution of the installation as well as it serves the precise scientific aim of grouping visitors in "batches" of 1 minute. At the beginning of every minute, an overhead structure made of LEDs lights up. The motion of visitors underneath influences the illumination patterns creating an interactive show. This brings visitors to stay within the "interactive experience" zone until when, about 50 seconds after the start, the show terminates in darkness (see [8] for further information).

Visitors, in a now almost dark environment, make then their way into the "measurement zone", the scientific core of the installation. Here, individuals choose between two symmetric exit ways. The exits are defined by a central obstacle which resembles, in shape, a liquid drop with squared sides. The obstacle measures about 5m in length, more than 3m in height, and 2m in width in its largest section – thus, it divides the corridor transversally into three even segments and the exits remain identified by its side ways.

As the flow is unidirectional and visitors have clear sight on the empty sidewalk past the installation, we expect a 50:50 choice rate between the exits, with possibly a slight preference for the right side due to cultural preferences (e.g. [9,10]). As mentioned, we apply stimuli to break the symmetry in the scenario aiming at influencing the route choice, which we measure by high-resolution individual tracking.

## 2.1 Arrow stimuli

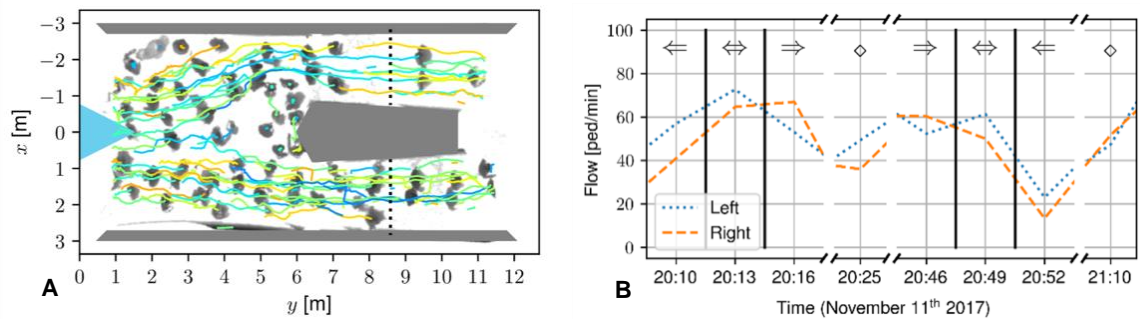

Figure 4: A depth-map collected by the Kinect-based measurement system with superimposed individual trajectories. Depth maps from the 12 sensors (cf. Figure 3A) are undistorted and merged following the procedure in [11]. The count line (coordinates $y = 8.5m$) used to evaluate instantaneous flows in Figures 5A and 7 is reported as dotted lines. Lateral fabric fences and central obstacle have been added manually in grey colour for reference. (B) Example measurements of pedestrian flows (ped./min) at the two sides of the obstacle as different stimuli (reported in the upper part of the plot) were applied. In the vast majority of cases, the left-right symmetry breaks with larger flow in the direction of the arrow as in the cases reported.

An arrow indication is possibly the simplest among the stimuli that can be employed to steer a crowd. Arrow indications were presented to the visitors through a squared LED matrix placed on the frontal face of the obstacle (size about 2m x 2m, cf. Figure 2). The presence of the arrow breaks the symmetry in the installation and comes with the obvious expectation for a quick preference shift for the side indicated. To check the effectiveness of the arrow stimuli we compare its effect against symmetric control conditions, specifically: an arrow pointing both ways and a stimuli-free condition where no arrow is displayed (see Figures 2CD).

## 2.2 Measurements

To quantify the effect of the stimuli we tracked automatically and with high-space-and-time-resolution each visitor crossing the measurement zone. To this aim we employed two independent technologies:
- S1. A state-of-the-art tracking system developed for pedestrian dynamics research. The system is based on a grid of overhead depth sensors (12 Microsoft Kinects™ [12], cf. Figure 3A) and on ad-hoc localization and tracking algorithms [6,11,13] (see also similar implementation in [14]).
- S2. A commercial pedestrian tracking system produced by Xovis AG, and here deployed in 5 overhead stereo cameras [15].

In Figures 3ABC we report, respectively, the sensors distribution, the number of visitors that crossed the installation each evening (inclusive of measurement differences between S1 and S2), and, as a sample, the time history of the pedestrian flow during the first evening.

Because the tracking systems S1 and S2 produced close results in the counts - especially in the conditions treated in this paper - and given the possibility of the system S1 of comparing the measurements directly with the recorded depth maps (cf. Figure 4A), in the following we will address only measurements by the system S1 (Kinect-based).

## 2.3 Steering ratio

We quantify the steering effect $S(j)$ of a stimulus $j$ as the ratio between the number of pedestrians that exited the installation by the left side (say $\#(L|j)$) and the total number (i.e. the sum of passages on the left and on the right, say $\#(L|j) + \#(R|j)$. The counting is restricted to time periods in which $j$ was active).

In formulas, this reads

$$S(j) = \frac{\#(L|j)}{\#(L|j) + \#(R|j)} \tag{1}$$

which, in a frequentist interpretation of probability, quantifies the probability of using the left exit conditioned to the presence of $j$. Operatively, we evaluate $S(j)$ by counting the crossing events of a transversal count-line (i.e. with coordinates $y = y_S = const$, cf. dashed line in Figure 4A). Passages with $x < 0$ identify a preference (or a choice in case of the position of the count-line in Figure 4A) for the left exit (and vice versa for the right case). An overall crowd-level preference for the left side thus yields $S(j) > 0.5$. Considering different count line positions (i.e. varying $y_s$) enables us to analyze the evolution of the side preference as a function of the distance to the obstacle.

### 3. Steering effect of arrow stimuli

During the festival, the four stimuli (left/right arrow, double arrow and no-arrow, cf. Figure 2) were activated in intervals of three minutes, respectively 61, 62, 60 and 52 times. Each one was thus active for about three hours, during which, a total of, respectively, 9017, 10060, 9895 and 8801 pedestrians crossed the facility.

As expected, we could generally observe a stabilization to a side preference following the arrow indication, or, for the control stimuli, to a roughly symmetric flow. In Figure 4B, we report, as an example, the readings of the pedestrian flow per minute as different stimuli were presented. Although in the figure the effect remain appreciable, instantaneous readings are generally noisy, as affected by the stochasticity of individual behaviours. A side preference, instead, robustly emerges as we consider the measurements as an ensemble,

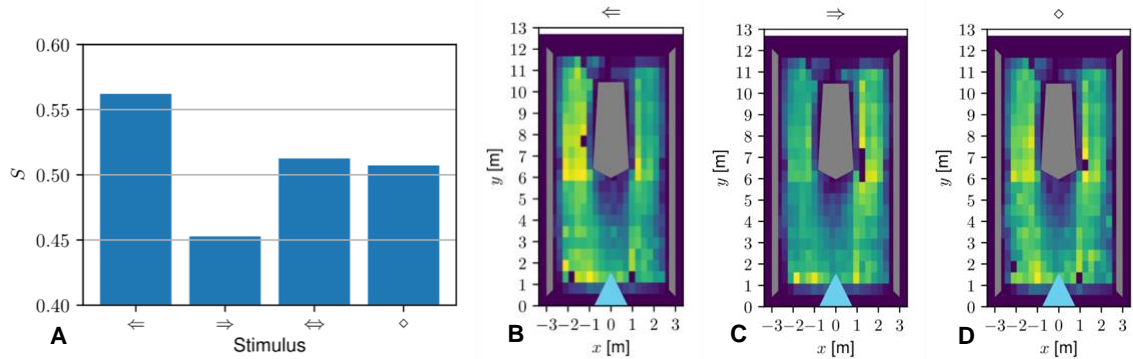

Figure 5: (A) Overall steering effect achieved by the considered stimuli (Figure 2) and quantified as the ratio $S(j)$ between passages by the left side and the total passages in presence of stimulus $j$ ($j \in \{\Leftarrow, \Rightarrow, \Leftrightarrow, \diamond\}$, cf. Equation 1). $S(j) > 0.5$ indicates a larger flow on the left side, and vice versa for the right side (cf. Section 2.3). These ratios encompass all flow and density conditions. The two arrow stimuli yield a nearly symmetric effect and varied the preference for the left side of about $\pm 6\%$ (in relation to the average neutral condition response). We notice further that for both neutral conditions, a preference of about $1\%$ for the left side is observed. (B-D) Probability distribution function of pedestrian positions in form of heatmaps, respectively in combination with a left, a right and a no-arrow stimulus. The routing preference reported in (A) is here noticeable as the side of the obstacle indicated by the arrow remains more used, than in the neutral case.

i.e. discarding the time variable, and compute the ratios $S(j)$ globally (see Figure 5A).

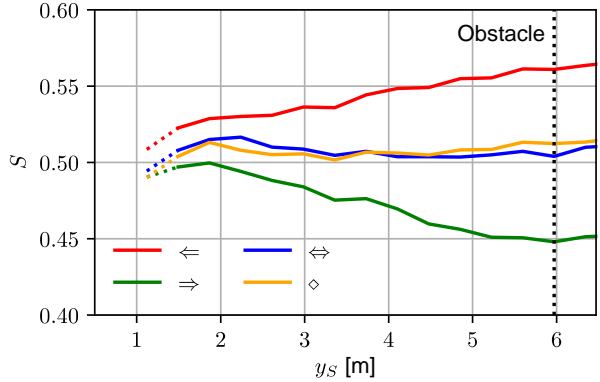

Figure 6: Steering ratio $S(j)$ (cf. Equation 1) as a function of the count line location $y_S$ (i.e. line $y = y_S$) which spans the interval between the entrance of the measurement region ($y \approx 0$; here we report the case $y_S > 1.5m$ as the trajectory reconstruction quality at the boundary of the measurement zone is lower) and the obstacle ($y \approx 6m$). At $y_S = 1.5m$, the steering ratio of arrow signs measures about $\pm 1.5\%$ (in comparison to control conditions). As we get closer to the obstacle, the steering ratio in case of arrows grows almost linearly up to $\pm 6\%$ (in comparison to control conditions) reaching the proportions displayed in Figure 5A. Extrapolating from the trend shown, we expect that around $y = 0$, no reaction to the stimuli is observable (i.e. $S(j) \approx S(control)$).

In the case of neutral stimuli, we observe a slight, yet surprising, preference for the left side (about, on average, 1% - i.e., $S(j) - 0.5 \approx 1\%$). This is necessarily connected with the environment: the rightmost fabric fence of the exhibit was about two meters far from a building, while the leftmost fence was bounding a large street (closed to the car traffic during the festival). Since these fabric fences were semi-transparent, we imagine, but we are not sure, that this could pass to the visitor impressions of higher prospect and/or of higher motion freedom on the left side thus the slightly higher preference. These ratios, in any case, define the baseline for our steering measurements and the reference with which we compare the steering ratios in case of arrow signs.

As we display arrow indications, routing choices, as expected, deviated from this baseline. We recorded an almost symmetric response for the left and right arrows in comparison with neutral conditions: in both cases the preference for the left incremented or, respectively, decremented by 6%. This effect remains qualitatively observable also in the position heatmaps in Figure 5BC, where the indicated side display higher position probability, i.e. most chosen in comparison with the neutral case in Figure 5D.

In Figure 6 we report the value of $S(j)$ as we move the count line position, parametrized by $y_S$, between the beginning of the measurement zone and the obstacle location. We notice that the side preference establishes within the measurement zone, in fact $S(j) \approx S(control)$ at the entrance of the zone ($y \approx 0$). In other words, despite the fact that the obstacle and the indication can be seen from farther away, it is only about when pedestrians enter the measurement zone that deviations from uniform left-right distributions, as to pass around the obstacle, establish. Furthermore, the side preference shows an almost linear growth as we approach the obstacle in presence of directional arrow stimuli.

We observe further a dependence of individuals' "steerability" on the entrance position. Steering effects are in fact stronger if a pedestrian faces the obstacle at the entrance (around $x = 0$ in Figure 4A). In these conditions, they do not have a straight way to the exits and a side movement is necessary. Considering only pedestrians entering in the central segment $x \in [-0.5, 0.5]m$, we measure steering ratios $S(\Leftarrow) \approx 63\%$ and $S(\Rightarrow) \approx 37\%$ (cf. Figure 7), considerably higher than in the global case (Figure 5A). On the opposite, pedestrians that entered more to the sides of the installation, i.e. close to one of the lateral fabric fences, changed their path to follow the arrow indication only with very small probability.

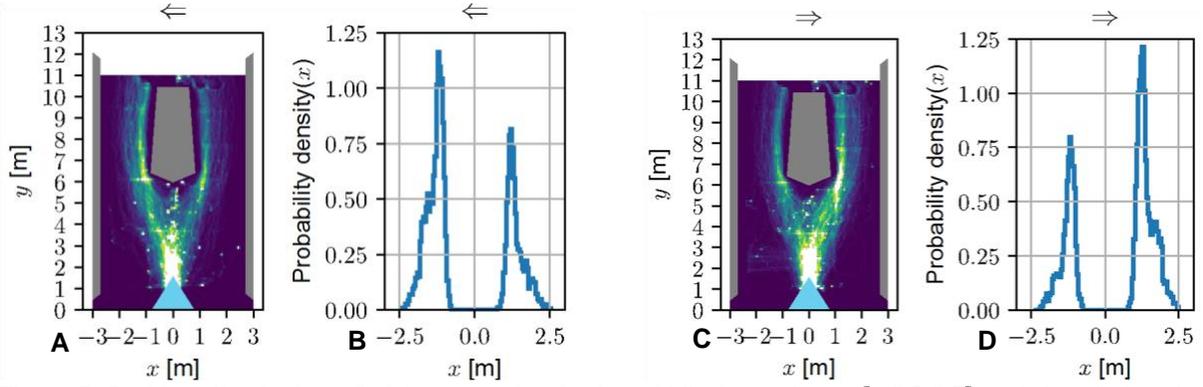

Figure 7: Position distribution of visitors entering in the middle interval $x \in [-0.5, 0.5]$ and thus facing the obstacle. (A, B) Case of left arrow stimulus, (C, D) case of right arrow stimulus. (A, C) position heatmap (as in Figures 5B-D). We observe that pedestrians remain close to the middle line and tend to stay close to the obstacle sides. (B, D) probability distribution function on counting line in Figure 4A; the corresponding steering ratios are $S(\Leftarrow) \approx 63\%$ and $S(\Rightarrow) \approx 37\%$. In these conditions our steering capability increases from 6% to 13%.

## 4. Discussion

In this paper, we introduced the "Moving Light" real-life experiment, in which we aimed at quantifying the effectiveness of visual stimuli in automatically steering crowd flows. In a week-long campaign held during Eindhoven Glow Festival 2017, we displayed stimuli to bias the decision of the over 140.000 visitors of the exhibit for one of the two otherwise symmetric exits. Our analysis of the stimuli effectiveness has been based on the exhaustive collection of all the visitors' trajectories through automatic high-resolution tracking methods. Gathering high volume of trajectories is a crucial aspect in our investigation and aims at ensuring high statistical resolution in the observations to build robust conclusions encompassing the randomness of individual behaviours.

We focused here on stimuli based on signage and, in particular, on arrow indications pointing toward one of the exits. Arrows are possibly the simplest stimulus one can devise to steer a choice between alternative directions. Despite the full freedom of choice, and the possibility to see past the installation, arrow indications generated higher pedestrian flows towards the indicated exit. Overall, we measured an increment of the flow of about 6% (with respect to the control conditions) in the direction pointed, and a symmetric response to the left and to the right arrows. Such a 6% increment generated within the measurement region (i.e. within $6m$ from the obstacle) and translates into having about 27% more people in the designated side with respect to the non-designated side (where $27\% = S(j)/(1 - S(j)) - 1 = (56\% - 44\%)/44\%$).

We notice that this ratio strongly increases as we just consider the individuals that face the obstacle when entering, and who thus cannot exit in a straight line; in this case $S(j) \approx 63\%$, and 70% more people pass by the designated side than by non-designated one. On the opposite, route choices of pedestrians entering along the sides of the installation are hardly affected by the arrow stimuli; instead, they tend to keep to the side on which they entered until the exit. We expect thus higher steering performance in corridors relatively smaller in width with respect to the obstacle. Besides, local density is likely to play a role in steering performance, which we will investigate in forthcoming studies.

We see the Moving Light experiment as a first step toward unmanned crowd steering devices leveraging on visual stimuli to influence individual route choice. This aims at solutions that enhance safety and comfort in civil infrastructures via optimized crowd routing. We established here an experimental benchmark for quantifying the effectiveness of steering stimuli, and we employed it to analyse the performance of arrow indications. As this paper is written, we are investigating the effects of the entire pool of stimuli considered in the Moving Light experiment. The results of these investigations will be reported on in future publications.

## Acknowledgements
This work is part of the JSTP research programme "Vision driven visitor behavior analysis and crowd management" with project number 341-10-001, which is financed by the Netherlands Organization for Scientific Research (NWO). We acknowledge the financial support of TU/e Intelligent Lighting Institute (ILI), 4TU and Philips Lighting, and the logistic support of Student Hotel Eindhoven. Moving Light has been possible also thanks to the contribution of M. Hoekstra, I. Iuncu, T. LeJeune, B. Maas, R. Nuij, S. Schippers, W. Willaert.